# Knightian Analysis of the Vickrey Mechanism[*]

Alessandro Chiesa
alexch@csail.mit.edu
MIT

Silvio Micali
silvio@csail.mit.edu
MIT

Zeyuan Allen Zhu
zeyuan@csail.mit.edu
MIT

April 24, 2015

**Abstract**

We analyze the Vickrey mechanism for auctions of multiple identical goods when the players have both Knightian uncertainty over their own valuations and incomplete preferences. In this model, the Vickrey mechanism is no longer dominant-strategy, and we prove that all dominant-strategy mechanisms are inadequate. However, we also prove that, in undominated strategies, the social welfare produced by the Vickrey mechanism in the worst case is not only very good, but also essentially optimal.

## 1 Introduction

We prove that the classical Vickrey mechanism guarantees good social welfare even when the players have *extremely* limited knowledge about themselves.

Recall that the Vickrey mechanism efficiently allocates multiple identical goods by ensuring that it is a dominant strategy for each player $i$ to report his true valuation, $\theta_i^*$. In real life, however, a player $i$ may be uncertain about $\theta_i^*$, as it may depend on variables that are not directly observable by him. A simple way to capture a player $i$'s uncertainty about his own valuation is the 'single-distribution' model, where $i$ does not know $\theta_i^*$, but only the true distribution from which $\theta_i^*$ has been drawn. We instead investigate a more general form of self uncertainty.

**Knightian Valuation Uncertainty.** In our model, the only information that a player $i$ has about $\theta_i^*$ (and more generally about the true valuation profile, $\theta^*$) consists of a *set of distributions*, from one of which $\theta_i^*$ has been drawn. We refer to this model as *Knightian valuation uncertainty* or *the Knightian valuation model*, as it is a special case of the uncertainty model envisaged by Frank H. Knight almost a century ago [19], and later formalized by Truman F. Bewley [4].

Knightian valuation uncertainty may arise from conflicting expert opinions. Consider a multi-unit auction of a novel good. Unable to evaluate his valuation, a player $i$ hires multiple (properly incentivized) independent experts to figure it out, trusting that at least one of them will be right. If each of them reports a different distribution for $\theta_i^*$, either because time was limited or because some of the experts made errors, then $i$ is ultimately faced with a set of distributions, from one of which $\theta_i^*$ has been drawn.

---

[*]We would like to thank the Econometrica editor and our anonymous referees for very valuable and patient guidance, without which these three computer scientists would have been unable to write the paper.

**Incomplete Preferences.** One may of course assume that players with Knightian valuation uncertainty have complete preferences, and in particular maxmin preferences, as defined by Gilboa and Schmeidler [16]. Such preferences are certainly defendable, however, quoting Bewley [4], they "do not lead to the sorts of economic behavior which make Knightian behavior interesting."

In our paper, players have *incomplete preferences*. A player $i$, only knowing that his true valuation has been selected from one of multiple distributions, prefers an outcome $\omega$ to another outcome $\omega'$ if and only if his expected utility for $\omega$ is higher than or equal to his expected utility for $\omega'$ with respect to all such distributions (and strictly greater for at least some of them). As a consequence, some outcomes or some strategies may be incomparable to him.

Finally, we do not assume that a player with incomparable strategies chooses a 'reference strategy'. That is, we do not rely on the *inertia* assumption of Bewley [4]. However, we assume that the players are risk-neutral.

**Findings.** In the Knightian valuation model, the Vickrey mechanism is no longer dominant-strategy, but multi-unit dominant-strategy mechanisms still exist: for instance, the 'degenerate' mechanism, which assigns all copies the good to a random player. Our Theorem 1 shows that all dominant-strategy mechanisms, as well as all ex-post Nash mechanisms, whether deterministic or randomized, must essentially be degenerate. That is, we provide natural conditions under which the allocations of such mechanisms are unresponsive to each player's action and thus cannot be efficient. Importantly, Theorem 1 applies also to mechanisms that allow a player to report a set of valuation distributions rather than a single valuation.

Since dominant-strategy mechanisms cannot achieve even an approximately efficient outcome in our model, it is natural to ask what social-welfare performance can be guaranteed in undominated strategies. After all, one may be quite confident that a player will not choose a strategy outside his undominated set.

Our Theorem 2 characterizes the set of undominated strategies of a player with Knightian valuation uncertainty in the Vickrey mechanism. A simple corollary of this characterization, Corollary 1, guarantees that, in undominated strategies, the social-welfare performance of the Vickrey mechanism is good even in the worst case.

This guarantee, of course, does not exclude that a different mechanism may perform even better. However, our Theorem 3 shows that the worst-case performance of the Vickrey mechanism is, *de facto*, asymptotically optimal among *all* undominated-strategy mechanisms, probabilistic or not, no matter what their strategy spaces may be. That is, as the number of players grows, no mechanism assigning finitely many pure strategies to each player can out-perform the Vickrey one in the worst case.

**In Sum.** Our theorems together show that, for risk-neutral players, the classical Vickrey mechanism is very robust to alternative specifications of preferences and information structures. Indeed, as most things classical, it outlives the confines in which it was conceived, and continues to be relevant in new and unforeseen settings. We believe that such robustness is an important property of a mechanism.

**Related Work.** Knightian uncertainty has received much attention in decision theory. Aumann [1]; Dubra, Maccheroni and Ok [11]; Ok [27]; and Nascimento [26] investigate decision with incomplete orders of preferences. Various criteria for selecting a single distribution out of a set of distributions have been studied by Danan [8]; Schmeidler [29]; Gilboa and Schmeidler [16]; and Maccheroni, Marinacci and Rustichini [23]. Bose, Ozdenoren and Pape [6] and Bodoh-Creed [5] use the model from Gilboa and Schmeidler [16] to study auctions. General equilibrium models with incompletely ordered preferences have been considered by Mas-Colell [24]; Gale and Mas-Colell [14];

Shafer and Sonnenschein [30]; and Fon and Otani [12]. Rigotti and Shannon [28] have characterized the set of equilibria in a financial market problem with incomplete preferences.

Mechanisms with Knightian uncertainty were first considered by Lopomo, Rigotti, and Shannon [21]. They do not focus on auctions, but on the rental extraction problem. (See Appendix G for a technical comparison.)

Lopomo, Rigotti, and Shannon also studied variants of the notions they proposed in [21] for a principal-agent model with Knightian uncertainty [22].

Di Tillio, Kos and Messner [10] and Bose and Renou [7] have studied *ambiguous* mechanisms, assuming that the players have maxmin preferences [16]. Informally, ambiguous mechanisms do not map a profile of strategies to a single outcome, but to an outcome arbitrarily chosen from a set of outcomes. Thus, in a sense, they 'exogenously introduce Knightian uncertainty'.

Full implementation in (traditional) undominated strategies was proposed by Jackson [17, 18]. An example of such implementation in the exact-valuation model is given by the mechanism of Babaioff et al. [3] for efficiency in multi-good auctions where each player may be interested in different bundles of the goods, but has the same value for each such bundle.

## 2 Model

### 2.1 Notation for Multi-Unit Auctions

We study auctions of a homogenous good in which players have multi-unit demand. We denote by $n$ the number of players; by $m$ the number of copies of the good; by $[n]$ the set $\{1, 2, \ldots, n\}$; and by $[m]$ the set $\{1, 2, \ldots, m\}$. The set of all possible allocations is $\mathcal{A} \stackrel{\text{def}}{=} \{A \in \mathbb{Z}_{\geq 0}^n \mid \sum_{i=0}^n A_i = m\}$. In an allocation $A \in \mathcal{A}$, $A_0$ is the number of unallocated copies and $A_i$ the number of copies allocated to player $i$.

As in [2, 31], we assume non-increasing marginal valuations. For each player $i$, the set of possible valuations is $\Theta_i \stackrel{\text{def}}{=} \{\theta_i : [m] \to \mathbb{R}_{\geq 0} \mid \theta_i(1) \geq \cdots \geq \theta_i(m) \geq 0\}$, where for each valuation $\theta_i \in \Theta_i$ and each copy $j \in [m]$, $\theta_i(j)$ represents player $i$'s marginal value for a $j$-th copy of the good. (We may also refer to such a $\theta_i$ as an $m$-dimensional vector, and to $\theta_i(j)$ as its $j$-th coordinate.) The set of all possible valuation profiles is $\Theta \stackrel{\text{def}}{=} \Theta_1 \times \cdots \times \Theta_n$. The profile of the players' true valuations is $\theta^* \stackrel{\text{def}}{=} (\theta_1^*, \ldots, \theta_n^*) \in \Theta$.

The set of possible outcomes is $\Omega \stackrel{\text{def}}{=} \mathcal{A} \times \mathbb{R}_{\geq 0}^n$. If $(A, P) \in \Omega$, we refer to $P_i$ as the price charged to player $i$. The utility of a player $i$, with valuation $\theta_i$, for an outcome $\omega = (A, P)$ is $U_i(\theta_i, \omega) \stackrel{\text{def}}{=} \sum_{j=1}^{A_i} \theta_i(j) - P_i$.

For every set $X$, we denote by $\Delta(X)$ the set of all countably additive probability measures on $X$. If $\omega \in \Delta(\Omega)$, then $U_i(\theta_i, \omega)$ is the expected utility of player $i$.

Relative to a valuation profile $\theta$, the social welfare of an outcome $\omega = (A, P) \in \Omega$, or the social welfare of an allocation $A \in \mathcal{A}$, is $\text{SW}(\theta, \omega) = \text{SW}(\theta, A) \stackrel{\text{def}}{=} \sum_i \sum_{j=1}^{A_i} \theta_i(j)$. The maximum social welfare relative to $\theta$ is $\text{MSW}(\theta) \stackrel{\text{def}}{=} \max_{A \in \mathcal{A}} \text{SW}(\theta, A)$. The maximum social welfare is $\text{MSW} \stackrel{\text{def}}{=} \text{MSW}(\theta^*)$.

A mechanism $M$ specifies, for each player $i$, a set of strategies $S_i$. We interchangeably refer to each member of $S_i$ as a pure *strategy/action/report* of $i$, and, similarly, to a member of $\Delta(S_i)$ as a mixed strategy/action/report of $i$.[1] After each player $i$, simultaneously with his opponents, reports a strategy $s_i$ in $S_i$, $M$ maps the reported strategy profile $s$ to an outcome $M(s) \in \Omega$. If $M$ is probabilistic, then $M(s) \in \Delta(\Omega)$.[2]

---

[1] Often, in pre-Bayesian settings, the notion of a strategy and that of an action are distinct. Indeed, a strategy $s_i$ of a player $i$ maps the set of all possible types of $i$ to the set of $i$'s possible actions/reports. But since strategies are universally quantified in all relevant definitions of this paper, we need not separate (and for simplicity do not separate) the notions of strategies and actions.

When in a mechanism $M$ the players jointly choose a profile of (possibly mixed) strategies $\sigma = (\sigma_1, \ldots, \sigma_n) \in \Delta(S_1) \times \cdots \times \Delta(S_n)$, we respectively denote by $M_i^{\mathsf{P}}(\sigma)$ and $M_{i,j}^{\mathsf{A}}(\sigma)$ the expected price of player $i$ and the probability that player $i$ wins $j$ copies of the good.

## 2.2 Knightian Valuation Uncertainty

In our model the players are risk-neutral and a player $i$'s sole information about the entire true valuation profile $\theta^* = (\theta_1^*, \ldots, \theta_n^*)$ consists of a non-empty set of distributions, $\mathcal{K}_i \subset \Delta(\Theta_i)$, from one of which $\theta_i^*$ has been drawn. (The players' true valuations are uncorrelated.)

Because a risk-neutral player cares only about his expected utility, and because in an auction each $\Theta_i$ is convex, in our model a player $i$ may 'collapse' each distribution $D_i \in \mathcal{K}_i$ to its expectation $\mathbb{E}_{\theta_i \sim D_i}[\theta_i] \in \Theta_i$. Accordingly, for auctions, our model can be equivalently restated in the following non-distributional language.

**Definition 2.1** (Knightian valuation model). *For each player $i$, $i$'s sole information about $\theta^*$ is a non-empty set $K_i \subset \Theta_i$, the candidate (valuation) set of $i$, such that $\theta_i^* \in K_i$. We refer to an element of $K_i$ as a candidate valuation. We denote by $\mathbb{K}_i$ the set of all possible candidate sets of $i$, and let $\mathbb{K} \stackrel{\text{def}}{=} \mathbb{K}_1 \times \cdots \times \mathbb{K}_n$.*

We stress that $\mathbb{K}_i$ can be an arbitrary subset of $2^{\Theta_i}$ and that, in our model, $i$ has no information about the true valuation $\theta_j^*$ or the candidate set $K_j$ of an opponent $j$.

In this paper, we refer to a player or an auction as *Knightian* to emphasize that we are considering the player or the auction in the Knightian valuation model.

In this model, a mechanism's performance will of course depend on the inaccuracy of the players' candidate sets, which we measure as follows.

**Definition 2.2.** *For all players $i$, candidate set $K_i$, and copies $j \in [m]$, we let*

$$K_i(j) \stackrel{\text{def}}{=} \{\theta_i(j) \mid \theta_i \in K_i\}, \quad K_i^{\perp}(j) \stackrel{\text{def}}{=} \inf K_i(j), \quad \text{and} \quad K_i^{\top}(j) \stackrel{\text{def}}{=} \sup K_i(j).$$

*A candidate set $K_i$ is (at most) $\delta$-approximate if $K_i^{\top}(j) - K_i^{\perp}(j) \leq \delta$ for all $j \in [m]$. An auction is (at most) $\delta$-approximate if, for each player $i$,*

$$\mathbb{K}_i \subset \mathbb{K}_i^{\delta} \stackrel{\text{def}}{=} \{K_i \in 2^{\Theta_i} \mid K_i \text{ is } \delta\text{-approximate}\}.$$

*We set $\mathbb{K}^{\delta} \stackrel{\text{def}}{=} \mathbb{K}_1^{\delta} \times \cdots \mathbb{K}_n^{\delta}$.*

Note that a candidate set $K_i$ may not be convex. For instance, in a single-good auction, $K_i$ may consist of the two valuations $a$ and $b$, and thus not contain $\frac{a+b}{2}$. Let us stress that the possibility of 'holes' in $K_i$ is the necessary sub-product of the fact that each $K_i$ is derived from an underlying set of distributions, $\mathcal{K}_i$, which is allowed to be totally arbitrary.[3]

---

[2]With our risk-neutral players, it would suffice to consider outcomes drawn from $\Delta(\mathcal{A}) \times \mathbb{R}_{\geq 0}^n$.

[3]Note that candidate sets may be very *expressive*. In a single-good setting, consider a player $i$ who believes that his true valuation is either $a$ or $b$, but more probably $a$ than $b$. This belief corresponds to the set of distributions $\mathcal{K}_i' = \{D_p \mid p \in [0.5, 1]\}$ where each $D_p$ is the distribution taking value $a$ with probability $p$, and value $b$ with probability $1 - p$. Then, if $i$ collapses each distribution $D_p$ to its expected value, he *de facto* ends with the following set of candidate valuations: $K_i' = \{pa + (1-p)b \mid p \in [0.5, 1]\} \subseteq \Theta_i$. (If, after translating the above belief to a new candidate set $K_i'$, player $i$ formed further beliefs about the probabilities of the valuations in $K_i'$, then he could again translate these beliefs to a new candidate set $K_i''$. And so on.)

When we say that the candidate set is $K_i$, we assume that all (partial) beliefs that player $i$ may have about his own valuation $\theta_i^*$ have already been taken into account.

# 3 The First Theorem

In this section, we prove that, under natural conditions, all dominant-strategy (and ex-post Nash) mechanisms must yield inefficient allocations in the Knightian valuation model. We stress that this result holds when such mechanisms are allowed to elicit from each player not just a single valuation, but an arbitrary report: in particular, a set of valuations.

Since it is easy to see that the revelation principle continues to apply in our setting (see Appendix A only for completeness sake), we state Theorem 1 in terms of Knightian dominant-strategy truthfulness mechanisms, formally defined below.

Recall that $\mathbb{K}_i$ is the set of all possible candidate sets of player $i$.

**Definition 3.1.** *A mechanism is* Knightian direct *if, for each player $i$, $S_i = \mathbb{K}_i$. Such a mechanism $M$ is* Knightian dominant-strategy-truthful (Knightian DST) *if*

$$\forall K_i, K_i' \in \mathbb{K}_i \ \forall K_{-i} \in \mathbb{K}_{-i} \ \forall \theta_i \in K_i \qquad U_i\big(\theta_i, M(K_i, K_{-i})\big) \geq U_i\big(\theta_i, M(K_i', K_{-i})\big).$$

To state Theorem 1, we also define a simple relation between candidate sets.

**Definition 3.2.** *In an $m$-unit auction, two candidate sets $K_i$ and $K_i'$ in $\mathbb{K}_i$ are*
- adjacent, *if* $\operatorname{span}\big\{\big(\theta_i(1) - \theta_i'(1), \ldots, \theta_i(m) - \theta_i'(m)\big) \,\big|\, \theta_i, \theta_i' \in K_i \cap K_i'\big\} = \mathbb{R}^m$, *and*
- connected, *if there exist $K_i^{(1)}, \ldots, K_i^{(t)} \in \mathbb{K}_i$ such that $K_i = K_i^{(1)}$, $K_i' = K_i^{(t)}$, and $K_i^{(k)}$ is adjacent to $K_i^{(k+1)}$ for all $k \in \{1, \ldots, t-1\}$.*

**Example 3.3.** When $m = 1$, that is, in the case of single-good auctions, each candidate set is a subset of the non-negative reals, and thus two candidate sets $K_i$ and $K_i'$ in $\mathbb{K}_i$ are adjacent if and only if $|K_i \cap K_i'| \geq 2$. Indeed, taking two different reals $x, y \in K_i \cap K_i'$, the fact that $x - y \neq 0$ implies that the 1-dimensional vector $(x - y)$ spans the 1-dimensional space $\mathbb{R}$. Accordingly, if the intervals $[1,3]$, $[2,4]$, and $[3,5]$ are possible candidate sets in $\mathbb{K}_i$, then $[1,3]$ is adjacent to $[2,4]$, $[2,4]$ is adjacent to $[3,5]$, and $[1,3]$ is connected (but not adjacent) to $[3,5]$.

Consider next an $m$-unit auction. Let $K_i$ be the candidate set consisting of all the valuations $\theta_i \in \Theta_i$ such that $\theta_i(j) \in [1,3]$ for all $j \in [m]$, and $K_i'$ the candidate set consisting of all the valuations $\theta_i' \in \Theta_i$ such that $\theta_i'(j) \in [2,4]$ for all $j \in [m]$. Then, $K_i$ and $K_i'$ are adjacent if they both belong to $\mathbb{K}_i$. This is so because the set of $m$-dimensional vectors $\big\{\big(\theta_i(1) - \theta_i'(1), \ldots, \theta_i(m) - \theta_i'(m)\big) \,\big|\, \theta_i, \theta_i' \in K_i \cap K_i'\big\}$ contains the $m$ vectors $(1, 0, \ldots, 0), (0, 1, 0, \ldots, 0), \cdots, (0, \ldots, 0, 1)$, which span $\mathbb{R}^m$.

**Theorem 1.** *In an $m$-unit Knightian auction, for all $\delta > 0$, all $\mathbb{K} \subseteq \mathbb{K}^\delta$, all (possibly probabilistic) Knightian DST mechanisms $M$,[4] all $(K_1, \ldots, K_n) \in \mathbb{K}$, all players $i$, all $K_i' \in \mathbb{K}_i$ connected to $K_i$, and all copies $j \in [m]$,*

$$M_{i,j}^{\mathsf{A}}(K_i, K_{-i}) = M_{i,j}^{\mathsf{A}}(K_i', K_{-i}) \quad and \quad M_i^{\mathsf{P}}(K_i, K_{-i}) = M_i^{\mathsf{P}}(K_i', K_{-i}) \ .$$

The proof of Theorem 1 can be found in Appendix B.

Theorem 1 essentially states that the probability that a Knightian DST mechanism $M$ assigns a given number of copies of the good to a given player $i$, and also the price player $i$ pays, are independent of the candidate sets $i$ reports, provided that they are connected and that the reports of $i$'s opponents are fixed.

---
[4] Note that Theorem 1 holds even if the mechanism $M$ is allowed to know $\delta$ and $\mathbb{K}$ in advance.

This independence from individual players' reports prevents a Knightian DST mechanism from guaranteeing high social welfare, when the players' possible candidate sets are sufficiently rich. For instance, consider a single-good auction in which $\delta = 2$, and each $\mathbb{K}_i$ includes the intervals $[0, 2]$, $[1, 3]$, $[2, 4], \ldots, [B, B + 2]$ for some large integer $B$. Then, no matter what the DST mechanism $M$ might be, when the reported profile of candidate sets is $K = ([0, 2], [0, 2], \ldots, [0, 2]) \in \mathbb{K}$, one of the players, without loss of generality player 1, must receive the good with probability at most $1/n$: in symbols, $M_{1,1}^{\mathsf{A}}(K) \leq 1/n$. This implies that the probability that player 1 gets the good remains at most $1/n$ even when all his opponents report the interval $[0, 2]$ and he reports $[B, B+2]$. This is so because the intervals $[0, 2]$ and $[B, B+2]$ are connected and thus Theorem 1 implies that $M_{1,1}^{\mathsf{A}}([0, 2], [0, 2], \ldots, [0, 2]) = M_{1,1}^{\mathsf{A}}([B, B + 2], [0, 2], \ldots, [0, 2])$. Accordingly, if $[B, B + 2]$ were the true candidate set of player 1, and $[0, 2]$ the true candidate set for everyone else, then the maximum social welfare would be at least $B$, while the expected social welfare delivered by $M$ would be at most $B/n + 2$.[5]

## 4 The Second Theorem

Our second theorem proves a very attractive relationship between a player's candidate set and his undominated strategies in the Vickrey mechanism for multi-unit Knightian auctions.

Recall that the Vickrey mechanism, denoted by Vickrey, is a direct mechanism (i.e., satisfies $S_i = \Theta_i$) and maps a profile of valuations $\theta \in \Theta_1 \times \cdots \times \Theta_n$, to an outcome $(A, P)$; where $A \in \arg\max_{A \in \mathcal{A}} \mathsf{SW}(\theta, A)$, $P_i = \mathsf{MSW}(\theta_{-i}) - \sum_{k \neq i} \sum_{j=1}^{A_k} \theta_k(j)$, and possible ties are broken lexicographically.[6]

For the Knightian valuation model, we define undominated strategies as follows.

**Definition 4.1.** *In a mechanism $M$, a pure strategy $s_i \in S_i$ of a player $i$ is* (weakly) dominated *by another possibly mixed strategy $\sigma_i \in \Delta(S_i)$ of $i$ with respect to his $K_i$, in symbols $s_i \prec_{(i, K_i)} \sigma_i$, if*

*(1) $\forall \theta_i \in K_i \ \forall s_{-i} \in S_{-i} \quad U_i(\theta_i, M(\sigma_i, s_{-i})) \geq U_i(\theta_i, M(s_i, s_{-i}))$, and*

*(2) $\exists \theta_i \in K_i \ \exists s_{-i} \in S_{-i} \quad U_i(\theta_i, M(\sigma_i, s_{-i})) > U_i(\theta_i, M(s_i, s_{-i}))$.*[7]

*A strategy $s_i \in S_i$ is* (weakly Knightian) undominated, *if there exists no $\sigma_i \in \Delta(S_i)$ such that $s_i \prec_{(i, K_i)} \sigma_i$. We denote the set of undominated strategies of player $i$ by $\mathsf{UD}_i(K_i)$.*

*If $K$ is a product or a profile of candidate sets, that is, if $K = (K_1, \ldots, K_n)$ or $K = K_1 \times \cdots \times K_n$, then $\mathsf{UD}(K) \stackrel{\mathrm{def}}{=} \mathsf{UD}_1(K_1) \times \cdots \times \mathsf{UD}_n(K_n)$.*

---

[5]We note that such poor social-welfare performance indeed relies on the richness of the players' possible candidate sets. If the players' possible candidate sets were guaranteed to be *sufficiently separated*, then a properly designed dominant-strategy mechanism could always achieve the *maximum* social welfare. For instance, consider an $n$-player auction of a single good where
- the inaccuracy parameter $\delta = 1/3$,
- the set of possible candidate sets $\mathbb{K}_i = \{[kn + i, kn + i + \frac{1}{3}] \mid k \in \mathbb{Z}_+\}$ for each player $i$, and
- the mechanism $M$ is such that (1) $S_i = \{kn + i \mid k \in \mathbb{Z}_+\}$ for each player $i$, and (2) for all $s \in S_1 \times \cdots \times S_n$, $M(s) = \mathsf{2P}(s)$, where $\mathsf{2P}$ is the second-price mechanism.

Then, it is clear that (a) for player $i$ whose true candidate set is $[kn + i, kn + i + \frac{1}{3}]$, reporting $kn + i$ is a dominant strategy, and (b) when dominant strategies are played, $M$ produces an outcome with maximum social welfare.

[6]More precisely, on a reported valuation profile $\theta$, the Vickrey mechanism sorts the values $\{\theta_i(j) \mid i \in [n], j \in [m]\}$ in a non-increasing order, and then chooses the $m$ largest entries to assign the $m$ copies of the good. Namely, if $\theta_i(1), \theta_i(2), \ldots, \theta_i(j)$ belong to the largest $m$ entries, but not $\theta_i(j + 1)$, then Vickrey assigns $j$ copies of the good to player $i$. If ties occur in this ordering, that is, if $\theta_i(j) = \theta_{i'}(j')$, then $\theta_i(j)$ precedes $\theta_{i'}(j')$ if and only if either (1) $i < i'$ or (2) $i = i'$ and $j < j'$.

[7]This notion is thus different from *strong dominance*, where inequality (1) is always strict. For strong dominance in the the exact-valuation case, see, for instance, [13, 20].

Our notion of an undominated strategy intends to capture the 'weakest condition' for which $s_i$ should be discarded in favor of $\sigma_i$, and is a natural extension of its classical counterpart.[8]

Note that Jackson's more involved definition of an undominated strategy is not necessary in our paper.[9]

Now let us formally state our second theorem.

> **Theorem 2.** *In an m-unit Knightian auction with the Vickrey mechanism, for all players $i$ and all candidate sets $K_i$, the set of undominated strategies $\mathsf{UD}_i(K_i)$ coincides with the set of all strategies $v_i \in \Theta_i$ satisfying the following condition*
> $$\forall j \in [m] \quad v_i(j) \in \left[K_i^\perp(j), K_i^\top(j)\right] \ .$$

Theorem 2 is proved in Appendix C.

Theorem 2 is obvious for $m = 1$, but less obvious when there are multiple copies of the good. In particular, a player $i$ may consider 'under-reporting' his value for the $j$-th copy of the good, but 'over-reporting' his value for the $k$-th copy. For example, in a 3-unit auction, where $K_i$ consists of all valuations $\theta_i \in \Theta_i$ such that

$$\theta_i(1) \in [100, 110], \quad \theta_i(2) \in [95, 105], \quad \text{and} \quad \theta_i(3) \in [90, 100] \ ,$$

by reporting the valuation $v_i = (113, 98, 80)$, $i$ over-reports his value for the first copy but under-reports that for the third copy. Such a strategy $v_i$ is, in general, *not dominated* by reporting the highest —respectively, the lowest— possible value for each copy of the good: that is, it is not dominated by reporting $(110, 105, 100)$ —respectively, $(100, 95, 90)$. However, one can still carefully construct a strategy $v_i^*$ that dominates $v_i$, and therefore conclude that a rational player will not use any such strategy $v_i$. In our example, such a $v_i^*$ could be $(110, 98, 80)$, $(113, 98, 90)$, or $(110, 98, 90)$. A general (but not the only) way to construct a $v_i^*$ dominating $v_i$ is to set $v_i^*(j) = v_i(j)$ for every copy $j$ such that $v_i(j)$ belongs to $[K_i^\perp(j), K_i^\top(j)]$, set $v_i^*(j) = K_i^\perp(j)$ if $v_i < K_i^\perp(j)$, and set $v_i^*(j) = K_i^\top(j)$ if $v_i > K_i^\top(j)$. (We rely on the non-increasing marginal valuation assumption in order to show that the so-constructed $v_i^*$ dominates $v_i$.)

The above construction of $v_i^*$ is the key idea to show that every $v_i \in \mathsf{UD}_i(K_i)$ satisfies $v_i(j) \in \left[K_i^\perp(j), K_i^\top(j)\right]$ for every copy $j$. A similar idea is needed to show the other direction. The details can be found in Appendix C.

**Remark 4.2.** For a Knightian player $i$, the set of undominated strategies $\mathsf{UD}_i(K_i)$ may strictly contain the candidate set $K_i$. For example, in a single-good second-price auction, if $K_i = \{4, 7, 21\}$, then not only reporting 4, 7, or 21 is an undominated strategy for player $i$, but so is reporting 9. As for another example, in a 2-unit Vickrey auction, if $K_i = \{(78, 60), (80, 50)\}$, then not only reporting $(78, 50)$ and $(80, 50)$, but also $(79, 51), (79, 52), \ldots, (79, 59)$, etc., are undominated strategies.

---

[8]Of course, other extensions are also possible. To express condition (2) in Definition 4.1, we must quantify the true valuation $\theta_i \in K_i$ and the pure strategy subprofile of $i$'s opponents $s_{-i} \in S_{-i}$. There are three alternatives to consider. Namely, (a) $\forall \theta_i \forall s_{-i}$, (b) $\exists \theta_i \forall s_{-i}$, and (c) $\forall \theta_i \exists s_{-i}$. Alternatives (a) and (b) do not yield the classical notion of (weak) dominance when $K_i$ is a singleton. Alternative (c) fails to capture the 'weakest condition' for which $s_i$ should be discarded in favor of $\sigma_i$. (Indeed, since $\sigma_i$ is already no worse than $s_i$, for player $i$ to discard strategy $s_i$ in favor of $\sigma_i$, it should suffice for $s_i$ to be strictly worse than $\sigma_i$ for a *single* possible valuation $\theta_i \in K_i$.)

[9]To meaningfully deal with the possibility of having an infinite sequence of pure strategies one dominating another, Jackson put forward, in the exact-valuation case, a more involved notion of an undominated strategy [17]. However, this more involved notion is unnecessary, even in the Knightian setting, for the class of *bounded* mechanisms. This class includes the Vickrey and all finite mechanisms, and thus all mechanisms analyzed in this paper in undominated strategies.

The multiplicity of undominated strategies in the two examples above emphasizes that our Knightian player $i$ has incomplete preferences. Assume for a moment that he had complete preferences: for instance, maxmin preferences. Then, his only undominated strategy (and thus his only dominant strategy) would consist of reporting 4 in the former example, and $(78, 52)$ in the latter one.[10]

Theorem 2 has a simple corollary (proved for completeness in Appendix D) about the social-welfare performance of the Vickrey mechanism.

**Corollary 1.** *In an $m$-unit Knightian auction, for all $\delta \geq 0$, all products $K$ of $\delta$-approximate candidate sets, all profiles $v \in \mathsf{UD}(K)$, and all $\theta \in K$*

$$\mathsf{SW}\big(\theta, \mathsf{Vickrey}(v)\big) \geq \mathsf{MSW}(\theta) - 2m\delta \ .$$

That is, the social welfare realized by the Vickrey mechanism is at most $2m\delta$ away for the maximum one, no matter which undominated strategies the players may choose. The following example shows that this performance guarantee of the Vickrey mechanism is actually tight in the worst case.

**Example 4.3.** Consider a two-player 10-approximate $m$-unit auction in which the candidate sets are

$$K_1 = \big\{(90, 90, \ldots, 90), (100, 100, \ldots, 100)\big\}$$
$$K_2 = \big\{(100, 100, \ldots, 100), (110, 110, \ldots, 110)\big\} \ .$$

In this case, the Vickrey mechanism may miss the maximum social welfare by $2\delta m$ as follows. Player 1 is 'optimistic' and bids the valuation $v_1 = (100, \ldots, 100)$; player 2 is 'pessimistic' and bids $v_2 = (100, \ldots, 100)$; the Vickrey mechanism (with the lexicographic tie-breaking rule) allocates all copies of the good to player 1; the true valuation $\theta_1^*$ of player 1 is $(90, \ldots, 90)$; and the true valuation $\theta_2^*$ of player 2 is $(110, \ldots, 110)$.

Accordingly, the realized social welfare is $90m$, while the maximum one is $110m = 90m + 2m\delta$. □

The relevance of worst-case analyses can of course be debated, but if the worst-case performance is good, then the typical performance can only be better. In our setting, a social-welfare loss of $2m\delta$ is small whenever $\delta$ is small relative to $\mathsf{MSW}(\theta)/m$. For instance, this is the case of a 10-unit auction in which a player's valuation for each copy of the good is a million dollars plus or minus \$100. Indeed, in this case, Corollary 1 implies that the Vickrey mechanism guarantees that the realized social welfare will be at least ten millions minus \$2,000, no matter how the players may choose their undominated strategies. (A performance loss that is at most linear in $\delta$ should not be underestimated. After all, Theorem 1 shows that the social welfare performance of any DST mechanism can be terrible, no matter how small, but positive, $\delta$ may be.)

---

[10] In the spirit of Gilboa and Schmeidler [16], given an outcome $\omega = (A, P)$, we can define the worst-case utility of a player $i$ with candidate set $K_i$ to be $\min_{\theta_i \in K_i} U_i(\theta_i, \omega) = \min_{\theta_i \in K_i} \sum_{j=1}^{A_i} \theta_i(j) - P_i$. In the above 2-unit auction example, this worst-case utility is $-P_i$, if $A_i = 0$; is $78 - P_i$, if $A_i = 1$; and is $130 - P_i$, if $A_i = 2$. Therefore, $i$'s worst-case utility coincides with the utility of a (non-Knightian) player whose true valuation is precisely $(78, 52)$. Indeed, $\min_{\theta_i \in K_i} U_i(\theta_i, \omega) = U_i\big((78, 52), \omega\big)$ for *all* possible outcomes $\omega$. Now, a player $i$ with maxmin preferences compares every two outcomes $\omega$ and $\omega'$ using his worst-case utility function, $\min_{\theta_i \in K_i} U_i(\theta_i, \cdot)$. It is thus equivalent for such a player $i$ to compare $\omega$ and $\omega'$ using the exact utility function $U_i\big((78, 52), \cdot\big)$. Accordingly, since the Vickrey mechanism is dominant-strategy in the classical setting, it is a dominant strategy for $i$ to report $(78, 52)$ in the above 2-unit auction.

# 5 The Third Theorem

Our third theorem shows that the worst-case social welfare performance of the Vickrey mechanism is essentially optimal, in the Knightian setting, relative to all possible undominated-strategy mechanisms.

Note that, in principle, there may be an undominated-strategy mechanism $M$ missing the maximum social welfare by at most $\delta m$.[11] Our upcoming Theorem 3, however, rules out the existence of such mechanisms, so long as they give each player a finite set of strategies.[12]

We stress that Theorem 3 applies not just to finite mechanisms eliciting a single valuation from each player, but to *all* finite mechanisms, including those allowing a player to report a set of valuations. Thus, in our Knightian valuation model, the social welfare optimality of the Vickrey mechanism (which allows a player to report only a single valuation) may be surprising.

We could simply state Theorem 3 by saying that, for every finite mechanism $M$, there exists a profile of $\delta$-approximate candidate sets for which $M$ misses the maximum social welfare by essentially $2m\delta$: more precisely, by $2m\delta(1 - 1/n) + \varepsilon$, where $\varepsilon$ is an arbitrarily small positive constant. To be more informative, however, we wish to state Theorem 3 so as to highlight the candidate set profiles causing this maximal loss in social welfare.

Let $V$ and $W$ be two sets of real numbers (with diameter at most $\delta$ and with at least two elements in common), whose union has diameter at least $2\delta - \varepsilon/m$. For instance, $V = [x - \delta, x]$ and $W = [x - 2\delta + \varepsilon/m, x - \delta + \varepsilon/m]$. Then, the following definition expresses that for each player there are at least two candidate sets, one for which the value of each copy of the good is in $V$, and one for which that the value of each copy of the good is in $W$. More precisely, recalling that $\mathbb{K}_i$ is the set of all possible candidate sets of player $i$, that $\mathbb{K} = (\mathbb{K}_1, \ldots, \mathbb{K}_n)$, and that in a $\delta$-approximate multi-unit Knightian auction $\mathbb{K} \subseteq \mathbb{K}^\delta$, we have the following

**Definition 5.1.** *In a $\delta$-approximate multi-unit Knightian auction, $\mathbb{K}$ is $\varepsilon$-basic if there exist two subsets $V$ and $W$ of non-negative numbers such that*
 *(a) $\max V - \min V \leq \delta$ and $\max W - \min W \leq \delta$,*
 *(b) $|V \cap W| > 1$ and $\max V - \min W \geq 2\delta - \varepsilon/m$, and*
 *(c) for every player $i$, $\mathbb{K}_i$ contains the following two candidate sets*

$$\widetilde{K}_i \stackrel{\text{def}}{=} \{\theta_i \in \Theta_i \mid \forall j, \theta_i(j) \in V\} \quad \text{and} \quad \widetilde{K}_i' \stackrel{\text{def}}{=} \{\theta_i \in \Theta_i \mid \forall j, \theta_i(j) \in W\} \ .$$

**Example 5.2.** Consider a 3-unit 10-approximate Knightian auction, in which for each player $i$, $\mathbb{K}_i$ includes the following two candidate sets:

$$[88, 98] \times [88, 98] \times [88, 98] \quad \text{and}$$
$$[80, 90] \times [80, 90] \times [80, 90] \ .$$

Then, $\mathbb{K}$ is 6-basic (corresponding to $V = [88, 98]$ and $W = [80, 90]$).

There is no magic about the choice of the numbers 80 and 88 in the above example. The main point is that the intersection of the two intervals $[80, 90]$ and $[88, 98]$ coincides with the interval $[88, 90]$, whose length is $\varepsilon/m = 6/3 = 2$.

---

[11] For instance, a mechanism $M$ could achieve such performance by asking each player $i$ to report a single valuation, and incentivizing him to report a valuation $v_i$ which is the 'mid-point' of his candidate set $K_i$: i.e., $v_i(j) = \frac{1}{2}(K_i^\perp(j) + K_i^\top(j))$ for all $j \in [m]$.

[12] This finiteness restriction, although crucial for our proof, is quite mild in practice (and is indeed natural when mechanisms are implemented via computers). The Vickrey mechanism itself becomes finite if it explicitly asks each player to report, for each copy of the good, an integral number of cents between 0 and $10^{100}$.

**Theorem 3.** *In a multi-unit Knightian auction, for all $\delta > 0$, all $\varepsilon > 0$, all $\varepsilon$-basic $\mathbb{K} \subseteq \mathbb{K}^\delta$, all (possibly probabilistic) finite mechanisms $M$, there exist products $K \in \mathbb{K}$, valuation profiles $\theta \in K$, and undominated strategy profiles $s \in \mathsf{UD}(K)$, such that*

$$\mathbb{E}\big[\mathrm{SW}(\theta, M(s))\big] \leq \mathrm{MSW}(\theta) - 2\delta m(1 - 1/n) + \varepsilon \ .$$

*Above, the expectation is over the possible random choices of the mechanism $M$.*

The proof of Theorem 3 can be found in Appendix E. Although mechanism finiteness is a natural restriction in practice, we wish to remark that Theorem 3 continues to hold under alternative but more complex assumptions.[13]

# Appendix

## A  Knightian Revelation Principle

Let us explicitly show that a version of the revelation principle [15, 9, 25] holds also in our Knightian setting. Recall that $\mathbb{K}_i$ is the set of all possible candidate sets for player $i$.

**Definition A.1.** *Let $M$ be a mechanism in which $S_i$ is the set of actions of player $i$. Then, the profile of functions $\big(s_i \colon \mathbb{K}_i \to \Delta(S_i)\big)_i$ is an* ex-post (possibly mixed) Nash equilibrium *of $M$ if for all $K \in \mathbb{K}$, all players $i$, all $a_i \in S_i$, and all $\theta_i \in K_i$,*

$$U_i(\theta_i, M(s_i(K_i), s_{-i}(K_{-i}))) \geq U_i(\theta_i, M(a_i, s_{-i}(K_{-i}))) \ .$$

**Lemma A.2** (**Revelation Principle**)**.** *Let $M$ be a mechanism that has an ex-post Nash equilibrium $s$. Then, there exists a Knightian DST mechanism $M'$ such that*

$$\forall K \in \mathbb{K} \quad M'(K_1, \ldots, K_n) = M(s_1(K_1), \ldots, s_n(K_1)) \ .$$

*Proof.* Let $M'$ be the Knightian direct mechanism so defined:

$$\forall K \in \mathbb{K} \quad M'(K_1, \ldots, K_n) \stackrel{\mathrm{def}}{=} M(s_1(K_1), \ldots, s_n(K_1)) \ .$$

(The above equality is between distributions if $M$ is probabilistic.)

All that is left to prove is that the mechanism $M'$ is dominant-strategy truthful. To this end, let $K_i$ be the true candidate set of player $i$. Then, for all $K'_i \in \mathbb{K}_i$, all $\theta_i \in K_i$, and all $K_{-i} \in \mathbb{K}_{-i}$,

$$U\big(\theta_i, M'(K_i, K_{-i})\big) = U\big(\theta_i, M(s_i(K_i), s_{-i}(K_{-i}))\big)$$
$$\geq U\big(\theta_i, M(s_i(K'_i), s_{-i}(K_{-i}))\big) = U\big(\theta_i, M'(K'_i, K_{-i})\big) \ ,$$

where the inequality follows from Definition A.1 of the ex-post Nash equilibrium by setting $a_i = s_i(K'_i)$. This completes the proof. □

Because every dominant-strategy mechanism must have an ex-post Nash equilibrium (consisting of each player choosing his dominant strategy), the above theorem holds also when $M$ is a dominant-strategy mechanism.

---

[13] As it will become clear from our proof, Theorem 3 also holds for all bounded mechanisms such that, for all players $i$, the strategy set $S_i$ is a compact Hausdorff space and, for all copies $j$, the families of allocation functions $\big\{M^{\mathsf{A}}_{i,j}(s_i, \cdot)\big\}_{s_i \in S_i}$ and price functions $\big\{M^{\mathsf{P}}_i(s_i, \cdot)\big\}_{s_i \in S_i}$ are equicontinuous. However, the Vickrey mechanism can be trivially modified to be finite, but not trivially made equicontinuous.

# B  Proof of Theorem 1

We start by proving, as a separate claim, that Theorem 1 holds in the case of *adjacent* (instead of *connected*) candidate sets. Namely,

**Claim B.1.** *For every player $i$, every two adjacent candidate sets $K_i, K_i' \in \mathbb{K}_i$ of $i$, and every subprofile $K_{-i}$ of candidate sets for $i$'s opponents,*

$$M_{i,j}^{\mathsf{A}}(K_i, K_{-i}) = M_{i,j}^{\mathsf{A}}(K_i', K_{-i}) \quad \text{and} \quad M_i^{\mathsf{P}}(K_i, K_{-i}) = M_i^{\mathsf{P}}(K_i', K_{-i}) \ .$$

*Proof.* Because the true candidate set of player $i$ may coincide with $K_i$, and because, when this is the case, reporting $K_i$ should dominate reporting $K_i'$, we have that, for all $\theta_i'' \in K_i$, the following inequality holds:

$$\sum_{j=1}^{m} M_{i,j}^{\mathsf{A}}(K_i, K_{-i}) \cdot \sum_{\ell=1}^{j} \theta_i''(\ell) - M_i^{\mathsf{P}}(K_i, K_{-i}) \geq \sum_{j=1}^{m} M_{i,j}^{\mathsf{A}}(K_i', K_{-i}) \cdot \sum_{\ell=1}^{j} \theta_i''(\ell) - M_i^{\mathsf{P}}(K_i', K_{-i}). \quad \text{(B.1)}$$

Similarly, because the true candidate set of player $i$ may coincide with $K_i'$, and because, when this is the case, reporting $K_i'$ should dominate reporting $K_i$, we also have that, for all $\theta_i'' \in K_i'$, the following inequality holds:

$$\sum_{j=1}^{m} M_{i,j}^{\mathsf{A}}(K_i', K_{-i}) \cdot \sum_{\ell=1}^{j} \theta_i''(\ell) - M_i^{\mathsf{P}}(K_i', K_{-i}) \geq \sum_{j=1}^{m} M_{i,j}^{\mathsf{A}}(K_i, K_{-i}) \cdot \sum_{\ell=1}^{j} \theta_i''(\ell) - M_i^{\mathsf{P}}(K_i, K_{-i}). \quad \text{(B.2)}$$

Next, for every pair of valuations $\theta_i, \theta_i' \in K_i \cap K_i'$, we choose $\theta_i'' = \theta_i$ in inequality (B.1) and $\theta_i'' = \theta_i'$ in inequality (B.2). Summing up the resulting inequalities, the $M_i^{\mathsf{P}}$ price terms cancel out, yielding the following inequality:

$$\sum_{j=1}^{m} M_{i,j}^{\mathsf{A}}(K_i, K_{-i}) \cdot \sum_{\ell=1}^{j} \big(\theta_i(\ell) - \theta_i'(\ell)\big) \geq \sum_{j=1}^{m} M_{i,j}^{\mathsf{A}}(K_i', K_{-i}) \cdot \sum_{\ell=1}^{j} \big(\theta_i(\ell) - \theta_i'(\ell)\big) \ . \quad \text{(B.3)}$$

Similarly, setting $\theta_i'' = \theta_i'$ in (B.1) and $\theta_i'' = \theta_i$ in (B.2), and summing up the resulting inequalities, we deduce that:

$$\sum_{j=1}^{m} M_{i,j}^{\mathsf{A}}(K_i, K_{-i}) \cdot \sum_{\ell=1}^{j} \big(\theta_i(\ell) - \theta_i'(\ell)\big) \leq \sum_{j=1}^{m} M_{i,j}^{\mathsf{A}}(K_i', K_{-i}) \cdot \sum_{\ell=1}^{j} \big(\theta_i(\ell) - \theta_i'(\ell)\big) \ . \quad \text{(B.4)}$$

(B.3) and (B.4) together imply, after some rearrangement of the terms, that

$$\sum_{j=1}^{m} \Big(M_{i,j}^{\mathsf{A}}(K_i, K_{-i}) - M_{i,j}^{\mathsf{A}}(K_i', K_{-i})\Big) \cdot \Big(\sum_{\ell=1}^{j} \theta_i(\ell) - \theta_i'(\ell)\Big) = 0 \ . \quad \text{(B.5)}$$

Now, let us denote by $a$ the $m$-dimensional vector such that $a_j \stackrel{\text{def}}{=} \sum_{k=j}^{m} M_{i,k}^{\mathsf{A}}(K_i, K_{-i}) - M_{i,k}^{\mathsf{A}}(K_i', K_{-i})$ for every $j \in [m]$. Then, (B.5) can be re-written as saying that the following inner product between two vectors is zero:

$$\big(a_1, \ldots, a_m\big) \cdot \big(\theta_i(1) - \theta_i'(1), \ldots, \theta_i(m) - \theta_i'(m)\big) = 0 \ . \quad \text{(B.6)}$$

Finally, using our assumption that the vectors $\big(\theta_i(1) - \theta_i'(1), \ldots, \theta_i(m) - \theta_i'(m)\big)$ span the entire $\mathbb{R}^m$, we easily conclude that $(a_1, \ldots, a_m) = (0, \ldots, 0)$, which in turn implies the desired equality of the allocation probabilities: $M_{i,j}^{\mathsf{A}}(K_i, K_{-i}) = M_{i,j}^{\mathsf{A}}(K_i', K_{-i})$. Plugging this equality into (B.1) and (B.2) immediately yields the desired equality of the prices: $M_i^{\mathsf{P}}(K_i, K_{-i}) = M_i^{\mathsf{P}}(K_i', K_{-i})$. □

It is now straightforward to see that Theorem 1 follows by the definition of connected candidate sets, and the repeated applications of the above claim. Namely, recall that $K_i, K_i' \in \mathbb{K}_i$ are connected if there exist $K_i^{(1)}, \ldots, K_i^{(t)} \in \mathbb{K}_i$ such that $K_i = K_i^{(1)}$, $K_i' = K_i^{(t)}$, and $K_i^{(k)}$ is adjacent to $K_i^{(k+1)}$ for all $k \in \{1, \ldots, t-1\}$. Therefore, we conclude that, for all $j \in [m]$,

$$M_{i,j}^{\mathsf{A}}(K_i^{(1)}, K_{-i}) = M_{i,j}^{\mathsf{A}}(K_i^{(2)}, K_{-i}) = \cdots = M_{i,j}^{\mathsf{A}}(K_i^{(t)}, K_{-i}) \;,$$

and similarly that

$$M_i^{\mathsf{P}}(K_i^{(1)}, K_{-i}) = M_i^{\mathsf{P}}(K_i^{(2)}, K_{-i}) = \cdots = M_i^{\mathsf{P}}(K_i^{(t)}, K_{-i}) \;. \qquad \blacksquare$$

## C  Proof of Theorem 2

Recall that the Vickrey mechanism is direct, that is, $S_i = \Theta_i$ for all players $i$. Recall also that multi-unit auctions have non-increasing marginal valuations, that is, $\theta_i(1) \geq \theta_i(2) \geq \cdots \geq \theta_i(m)$ for each $\theta_i \in \Theta_i$. Therefore, $K_i^\perp(1), \ldots, K_i^\perp(m)$ and $K_i^\top(1), \ldots, K_i^\top(m)$ are non-decreasing sequences. That is, $K_i^\top, K_i^\perp \in \Theta_i$. Accordingly, both $K_i^\top$ and $K_i^\perp$ are valid reports for player $i$ in the Vickrey mechanism.

We start by proving, by contradiction, that

$$v_i \in \mathsf{UD}_i(K_i) \implies v_i(j) \geq K_i^\perp(j) \text{ for all } j \in [m]. \tag{C.1}$$

Assume that implication (C.1) is false; let $j^* \in [m]$ be the first coordinate $j$ such that $v_i(j) < K_i^\perp(j)$; and define the function $v_i^* : [m] \to \mathbb{R}_{\geq 0}$ as follows:

$$v_i^*(j) = \begin{cases} v_i(j), & \text{if } j \neq j^*; \\ K_i^\perp(j), & \text{if } j = j^*. \end{cases}$$

Since $v_i$ and $K_i^\perp$ are monotonically non-increasing, so is $v_i^*$. Indeed,

- if $j^* > 1$, then $v_i^*(j^* - 1) = v_i(j^* - 1) \geq K_i^\perp(j^* - 1) \geq K_i^\perp(j^*) = v_i^*(j^*)$

- if $j^* < m$, then $v_i^*(j^*) = K_i^\perp(j^*) > v_i(j^*) \geq v_i(j^* + 1) = v_i^*(j^* + 1)$ .

Thus also $v_i^*$ is a valid valuation in $\Theta_i$. We now reach a contradiction by showing that $v_i^*$ weakly dominates $v_i$, that is,

$$\forall\, \theta_i \in K_i \; \forall\, v_{-i} \qquad U_i\big(\theta_i, \mathsf{Vickrey}(v_i^*, v_{-i})\big) \geq U_i\big(\theta_i, \mathsf{Vickrey}(v_i, v_{-i})\big) \;, \tag{C.2}$$

$$\exists\, \theta_i' \in K_i \; \exists\, v_{-i}' \qquad U_i\big(\theta_i', \mathsf{Vickrey}(v_i^*, v_{-i}')\big) > U_i\big(\theta_i', \mathsf{Vickrey}(v_i, v_{-i}')\big) \;. \tag{C.3}$$

To show (C.2), choose arbitrarily $v_{-i} \in \Theta_{-i}$, and consider the following two cases:

(1) *In $\mathsf{Vickrey}(v_i^*, v_{-i})$ and $\mathsf{Vickrey}(v_i, v_{-i})$, $i$ receives the same number of copies.*

   In this case, inequality (C.2) holds because its two sides are equal for all $\theta_i$.

(2) *In $\mathsf{Vickrey}(v_i^*, v_{-i})$ and $\mathsf{Vickrey}(v_i, v_{-i})$, $i$ receives different numbers of copies.*

   In this case, one can carefully verify that player $i$ wins $j^*$ copies of the good in $\mathsf{Vickrey}(v_i^*, v_{-i})$ and only $j^* - 1$ copies in $\mathsf{Vickrey}(v_i, v_{-i})$.[14] Thus, (C.2) holds because of the following two reasons:

---

[14] Recall that, when each player reports a valuation $v_i$, the Vickrey mechanism orders the $nm$ values $\{v_i(j) \mid i \in [n], j \in [m]\}$ (breaking ties lexicographically), and allocates the $m$ copies of the good by looking at the first $m$ values in this order. Since the only difference between $v_i^*$ and $v_i$ is that $v_i^*(j^*) > v_i(j^*)$, the ordering of the reported $nm$ values is minimally affected. That is, if player $i$ receives different numbers of copies in outcome $(v_i, v_{-i})$ and outcome $(v_i^*, v_{-i})$, then it must be that $v_i(j^*)$ is outside the largest $m$ numbers under $(v_i, v_{-i})$, but $v_i^*(j^*)$ is within the largest $m$ numbers under $(v_i^*, v_{-i})$. This implies that $i$ wins $j^* - 1$ copies in $\mathsf{Vickrey}(v_i, v_{-i})$ but $j^*$ copies in $\mathsf{Vickrey}(v_i^*, v_{-i})$.

- *i*'s price for his extra $j^*$-th copy of the good is $\leq K_i^\perp(j^*)$.
  Indeed, Vickrey guarantees that $i$ pays for his $j^*$-th copy at most the value he reports for it. That is, for his $j^*$-th copy, $i$ pays at most $v_i^*(j^*)$, which in turn is equal to $K_i^\perp(j^*)$.
- *i*'s value for this $j^*$-th copy is $\geq K_i^\perp(j^*)$.
  Indeed, for any candidate valuation $\theta_i$ in $K_i$, $\theta_i(j^*) \geq K_i^\perp(j^*)$.

Therefore, inequality (C.2) holds. Let us now show that also inequality (C.3) holds. To do so, we need to construct a 'witness' candidate valuation $\theta_i' \in K_i$ and a 'witness' strategy sub-profile $v_{-i}'$. In fact, we construct some $v_{-i}'$ so that (C.3) holds for all $\theta_i'$. Let $v_{-i}'$ be the strategy subprofile in which, for every player $k \neq i$,

$$\forall j \in [m] \qquad v_k'(j) \stackrel{\text{def}}{=} x \stackrel{\text{def}}{=} \frac{v_i(j^*) + K_i^\perp(j^*)}{2} < K_i^\perp(j^*) \ .$$

Then, player $i$ wins exactly $j^*$ copies in $\mathsf{Vickrey}(v_i^*, v_{-i}')$ and pays $x$ for each one of them; and wins exactly $j^* - 1$ copies in $\mathsf{Vickrey}(v_i, v_{-i}')$ and pays $x$ for each one of them. Indeed, there are exactly $j^* - 1$ numbers greater than $x$ in $v_i$, exactly $j^*$ in $v_i^*$, and $x$ is the reported value of every other player, in $v_{-i}'$, for every single copy of the good. As a result, $i$'s utility in $\mathsf{Vickrey}(v_i^*, v_{-i}')$ is strictly greater than that in $\mathsf{Vickrey}(v_i, v_{-i}')$. This is so because in the outcome $\mathsf{Vickrey}(v_i^*, v_{-i}')$, $i$ pays an extra price $x$ for his $j^*$-th copy, while being guaranteed that his true valuation for the $j^*$-th copy, $\theta_i(j^*)$, is strictly larger than $x$, because $x < K_i^\perp(j^*) \leq \theta_i(j^*)$. Therefore, we conclude that (C.3) holds for all candidate $\theta_i' \in K_i$ and the above defined $v_{-i}'$.

Since both (C.2) and (C.3) hold, valuation $v_i^*$ (weakly) dominates $v_i$, contradicting the hypothesis that $v_i \in \mathsf{UD}_i(K_i)$. This contradiction proves (C.1).

An absolutely symmetrical argument shows that[15]

$$v_i \in \mathsf{UD}_i(K_i) \implies v_i(j) \leq K_i^\top(j) \text{ for all } j \in [m]. \tag{C.4}$$

Together, statements (C.1) and (C.4) imply that all undominated strategies $v_i \in \mathsf{UD}_i(K_i)$ satisfy $v_i(j) \in [K_i^\perp(j), K_i^\top(j)]$ for each copy $j$.

Let us now prove the other direction: namely, that every strategy $v_i$ satisfying $v_i(j) \in [K_i^\perp(j), K_i^\top(j)]$ for each copy $j$ is undominated for player $i$.

We proceed by contradiction. Suppose that there exists a valuation $v_i^*$ that weakly dominates $v_i$. We are going to derive a contradiction by showing that

$$\exists \theta_i' \in K_i \ \exists v_{-i}' \qquad U_i\big(\theta_i, \mathsf{Vickrey}(v_i^*, v_{-i}')\big) < U_i\big(\theta_i, \mathsf{Vickrey}(v_i, v_{-i}')\big) \ . \tag{C.5}$$

Since, by the definition of weak dominance, $v_i^*$ must be different from $v_i$, there are two (not mutually exclusive) cases to consider:

(a) $v_i^*(j) > v_i(j)$ for some $j \in [m]$, and

(b) $v_i^*(j) < v_i(j)$ for some $j \in [m]$.

In case (a), let

- $j^*$ be the first coordinate $j \in [m]$ such that $v_i^*(j) > v_i(j)$,
- $y$ be a real number such that $v_i^*(j^*) > y > v_i(j^*)$, and
- $\varepsilon$ be a real number in the open interval $\big(0, y - v_i(j^*)\big)$.

To show (C.5), we let $\theta_i'$ be an arbitrary valuation in $K_i$ satisfying $\theta_i'(j^*) \leq v_i(j^*) + \varepsilon$. (Such a valuation always exists since $K_i^\perp(j^*) = \inf\{\theta_i(j^*) \,|\, \theta_i \in K_i\} \leq v_i(j^*)$.) Next, we construct the required strategy sub-profile $v_{-i}'$ as follows: for each player $k \neq i$ and each copy $j$, $v_k'(j) \stackrel{\text{def}}{=} y$. Let us now compare player $i$'s utilities in the outcomes $\mathsf{Vickrey}(v_i^*, v_{-i}')$ and $\mathsf{Vickrey}(v_i, v_{-i}')$.

---

[15] In this symmetrical case, one needs to define $j^* \in [m]$ to be the *last* coordinate such that $v_i(j^*) > K_i^\top(j)$.

In Vickrey$(v_i^*, v_{-i}')$, $i$ wins at least $j^*$ copies, because $v_i^*(1) \geq \cdots \geq v_i^*(j^*) > y$; moreover, he pays $y$ for each such copy, because $y$ is the value that every other player reports, in $v_{-i}'$, for every single copy of the good. By contrast, in Vickrey$(v_i, v_{-i}')$, $i$ wins exactly $j^* - 1$ copies, because $v_i(1) \geq \cdots \geq v_i(j^* - 1) \geq v_i^*(j^* - 1) > y$ and $v_i(j^*) < y$; moreover, he again pays $y$ for each of them. Thus, to prove that

$$U_i\big(\theta_i, \mathsf{Vickrey}(v_i^*, v_{-i}')\big) < U_i\big(\theta_i, \mathsf{Vickrey}(v_i, v_{-i}')\big) \ ,$$

it suffices to point out that, for each copy $j \geq j^*$ that $i$ wins in Vickrey$(v_i^*, v_{-i}')$, $i$'s true value is $\theta_i'(j) \leq \theta_i'(j^*) \leq v_i(j^*) + \varepsilon < y$. This ends the proof of (C.5) in case (a).

In case (b), we instead let $j^*$ be the last coordinate $j \in [m]$ such that $v_i^*(j) < v_i(j)$. An absolutely symmetrical argument shows that (C.5) also holds for this case.

In sum, Theorem 2 holds. ∎

## D  Proof of Corollary 1

Let $v \in \mathsf{UD}(K)$ be any profile of undominated strategies, and $A = (A_0, A_1, \ldots, A_n)$ represent the allocation in the outcome Vickrey$(v)$, where each player $i$ receives $A_i$ copies of the goods, and $A_0$ is the number of unallocated copies. For any $\theta \in K$, let $B = (B_0, B_1, \ldots, B_n)$ represent the allocation that maximizes social welfare under $\theta$, i.e., $B = \arg\max_{B \in \mathcal{A}} \big\{ \sum_{i=1}^n \sum_{\ell=1}^{B_i} \theta_i(\ell) \big\}$. Then,

$$\mathsf{SW}(\theta, \mathsf{Vickrey}(v)) = \sum_{i=1}^n \sum_{\ell=1}^{A_i} \theta_i(\ell) \stackrel{(1)}{\geq} \sum_{i=1}^n \sum_{\ell=1}^{A_i} \big(K_i^\top(\ell) - \delta\big) \stackrel{(2)}{\geq} \sum_{i=1}^n \sum_{\ell=1}^{A_i} \big(v_i(\ell) - \delta\big)$$

$$\stackrel{(3)}{\geq} \Big(\sum_{i=1}^n \sum_{\ell=1}^{A_i} v_i(\ell)\Big) - m\delta \stackrel{(4)}{\geq} \Big(\sum_{i=1}^n \sum_{\ell=1}^{B_i} v_i(\ell)\Big) - m\delta \stackrel{(5)}{\geq} \Big(\sum_{i=1}^n \sum_{\ell=1}^{B_i} K_i^\bot(\ell)\Big) - m\delta$$

$$\stackrel{(6)}{\geq} \Big(\sum_{i=1}^n \sum_{\ell=1}^{B_i} (\theta_i(\ell) - \delta)\Big) - m\delta \stackrel{(7)}{\geq} \mathsf{MSW}(\theta) - 2m\delta.$$

Above
- Inequality (1) holds because $\theta \in K$, and thus $\theta_i(\ell) \geq K_i^\bot(\ell) \geq K_i^\top(\ell) - \delta$;
- Inequality (2) holds by Theorem 2;
- Inequality (3) holds because we have only $m$ copies of the good: $\sum_{i=1}^n A_i \leq m$;
- Inequality (4) holds because the Vickrey mechanism maximizes social welfare with respect to $v$, and thus, relative to $v$, $(A_0, \ldots, A_n)$ is no worse than any other allocation, and in particular no worse than $(B_0, \ldots, B_n)$;
- Inequality (5) holds again by Theorem 2;
- Inequality (6) holds because $\theta \in K$, and thus $\theta_i(\ell) \leq K_i^\top(\ell) \leq K_i^\bot(\ell) + \delta$; and
- Inequality (7) holds because of the definition of $\mathsf{MSW}(\theta)$ and the fact that $\sum_{i=1}^n B_i \leq m$. ∎

## E  Proof of Theorem 3

### E.1  A Structural Lemma

The following lemma applies to *all* finite mechanisms, including those that allow players to report sets of valuations, or anything else. (Indeed, the revelation principle no longer holds for mechanisms that are not dominant-strategy or ex-post Nash. Thus, we must be able to deal with general mechanisms with arbitrary strategy spaces.)

**Lemma E.1.** *Let $M$ be a finite mechanism and $i$ a player, let $x = (x_1, \ldots, x_m)$ and $y = (y_1, \ldots, y_m)$ be two valuations in $\Theta_i$ such that $x_j > y_j$ for all copies $j \in [m]$, and let $K_i$ and $\widetilde{K}_i$ be two candidate sets for $i$ such that,*

$$\forall\, t \in \{0, 1, \ldots, m\} \quad (x_1, \ldots, x_t, y_{t+1}, \ldots, y_m) \in K_i \cap \widetilde{K}_i.^{16} \tag{E.1}$$

*Then, for every $\varepsilon > 0$, there are mixed strategies $\sigma_i \in \Delta(\mathsf{UD}_i(K_i))$ and $\widetilde{\sigma}_i \in \Delta(\mathsf{UD}_i(\widetilde{K}_i))$ such that, for all $s_{-i} \in S_{-i}$ and all $j \in [m]$,*

$$\left| M_{i,j}^{\mathsf{A}}(\sigma_i, s_{-i}) - M_{i,j}^{\mathsf{A}}(\widetilde{\sigma}_i, s_{-i}) \right| < \varepsilon \ .^{17}$$

*Proof.* First of all, it is simple to see (but anyway proved in Appendix F) that for every finite mechanism, the set of undominated strategies of a Knightian player is always non-empty. Therefore, the sets $\mathsf{UD}_i(K_i)$ and $\mathsf{UD}_i(\widetilde{K}_i)$ are both non-empty. If there exists a common (pure) strategy $s_i \in \mathsf{UD}_i(K_i) \cap \mathsf{UD}_i(\widetilde{K}_i)$, then setting $\sigma_i = \widetilde{\sigma}_i = s_i$ proves Lemma E.1. Therefore, let us assume in the rest of the proof that $\mathsf{UD}_i(K_i)$ and $\mathsf{UD}_i(\widetilde{K}_i)$ are *totally disjoint*.

Let $s_i$ be a pure strategy in $\mathsf{UD}_i(K_i)$. Then, $\mathsf{UD}_i(K_i) \cap \mathsf{UD}_i(\widetilde{K}_i) = \varnothing$ implies that $s_i \notin \mathsf{UD}_i(\widetilde{K}_i)$. By definition, $s_i \notin \mathsf{UD}_i(\widetilde{K}_i)$ implies the existence of a (possibly mixed) strategy $\widetilde{\sigma}_i \in \Delta(\mathsf{UD}_i(\widetilde{K}_i))$ that (weakly) dominates $s_i$ for player $i$ with respect to candidate set $\widetilde{K}_i$. In symbols, as per Definition 4.1, $\widetilde{\sigma}_i \succ_{(i, \widetilde{K}_i)} s_i$.

Next, we argue that

$$\exists\, \tau_i \in \Delta(\mathsf{UD}_i(K_i)) \text{ such that } \tau_i \succ_{(i, K_i)} \widetilde{\sigma}_i \ .^{18} \tag{E.2}$$

Let us write the possibly mixed strategy $\widetilde{\sigma}_i$ as a sum of pure ones, $\widetilde{\sigma}_i = \sum_{t \in X} \alpha^{(t)} \widetilde{s}_i^{(t)}$. Here, $X$ is a finite index set, each $\widetilde{s}_i^{(t)}$ is a pure strategy from $\mathsf{UD}_i(\widetilde{K}_i)$, each $\alpha^{(t)} > 0$, and $\sum_{t \in X} \alpha^{(t)} = 1$. Invoking again the disjointedness of $\mathsf{UD}_i(K_i)$ and $\mathsf{UD}_i(\widetilde{K}_i)$, we deduce that $\widetilde{s}_i^{(t)} \notin \mathsf{UD}_i(K_i)$ for each $t \in X$. This implies the existence of a strategy $\tau_i^{(t)} \in \Delta(\mathsf{UD}_i(K_i))$ such that $\tau_i^{(t)} \succ_{(i, K_i)} \widetilde{s}_i^{(t)}$. Thus, by defining $\tau_i \stackrel{\text{def}}{=} \sum_{t \in X} \alpha^{(t)} \tau_i^{(t)}$, we have that $\tau_i$ dominates $\widetilde{\sigma}_i$. Thus, (E.2) holds.

Similarly, we could argue that there exists some $\widetilde{\tau}_i \in \Delta(\mathsf{UD}_i(\widetilde{K}_i))$ such that $\widetilde{\tau}_i \succ_{(i, \widetilde{K}_i)} \tau_i$. Continuing in this fashion, going back and forth between $\Delta(\mathsf{UD}_i(K_i))$ and $\Delta(\mathsf{UD}_i(\widetilde{K}_i))$, we obtain an infinite chain of (possibly repeating) strategies,

$$\sigma_i^{(1)} \prec_{(i, \widetilde{K}_i)} \widetilde{\sigma}_i^{(1)} \prec_{(i, K_i)} \sigma_i^{(2)} \prec_{(i, \widetilde{K}_i)} \widetilde{\sigma}_i^{(2)} \prec_{(i, K_i)} \cdots$$

This (weak) dominance chain implies the following utility inequalities: for all $s_{-i} \in S_{-i}$ and all $k \in \mathbb{N}$:

$$\begin{array}{rl}
\forall \widetilde{\theta}_i \in \widetilde{K}_i & U_i\big(\widetilde{\theta}_i, M(\sigma_i^{(k)}, s_{-i})\big) \leq U_i\big(\widetilde{\theta}_i, M(\widetilde{\sigma}_i^{(k)}, s_{-i})\big) \\
\forall \theta_i \in K_i & U_i\big(\theta_i, M(\widetilde{\sigma}_i^{(k)}, s_{-i})\big) \leq U_i\big(\theta_i, M(\sigma_i^{(k+1)}, s_{-i})\big)
\end{array} \tag{E.3}$$

Next, for every $t \in \{0, 1, \ldots, m\}$, we define

$$z_t \stackrel{\text{def}}{=} (z_{t,1}, z_{t,2}, \ldots, z_{t,m}) \stackrel{\text{def}}{=} (x_1, x_2, \ldots, x_t, y_{t+1}, \ldots, y_m) \in K_i \cap \widetilde{K}_i \ .$$

Choosing $\theta_i = \widetilde{\theta}_i = z_t$ in (E.3), we obtain that for all $s_{-i} \in S_{-i}$ and all $k \in \mathbb{N}$,

$$U_i\big(\widetilde{\theta}_i, M(\sigma_i^{(k)}, s_{-i})\big) \leq U_i\big(\widetilde{\theta}_i, M(\widetilde{\sigma}_i^{(k)}, s_{-i})\big) = U_i\big(\theta_i, M(\widetilde{\sigma}_i^{(k)}, s_{-i})\big) \leq U_i\big(\theta_i, M(\sigma_i^{(k+1)}, s_{-i})\big) \ .$$

---

[16] Recall that all valuations in $\Theta_i$ are non-increasing. Our chosen vectors $(x_1, \ldots, x_t, y_{t+1}, \ldots, y_m)$ are indeed non-increasing, because we have $x_j > y_j$ and both $x$ and $y$ are non-increasing.

[17] In fact, Lemma E.1 can be strengthened to ensure that the prices are close too: namely, $\big|M_i^{\mathsf{P}}(\sigma_i, s_{-i}) - M_i^{\mathsf{P}}(\widetilde{\sigma}_i, s_{-i})\big| < \varepsilon$. However, this strengthened version of Lemma E.1 is not needed in order to prove Theorem 3.

[18] Note that, while we have only defined what it means for a *pure* strategy to be dominated by a possibly mixed one, the definition trivially extends to the case of dominated strategies that are *mixed*, as is the case in "$\tau_i \succ_{(i, K_i)} \widetilde{\sigma}_i$" in (E.2).

Putting together the above inequalities for $k = 1, 2, \ldots$, we get the following infinite and non-decreasing sequence of real numbers (for each $s_{-i} \in S_{-i}$):

$$U_i\big(z_t, M(\sigma_i^{(1)}, s_{-i})\big) \leq U_i\big(z_t, M(\widetilde{\sigma}_i^{(1)}, s_{-i})\big) \leq U_i\big(z_t, M(\sigma_i^{(2)}, s_{-i})\big) \leq \cdots$$

This sequence is upperbounded by $x_1 + \cdots + x_m$. (Indeed, $z_{t,l} \leq x_l$ for each $l$. So, the $i$'s valuation is at most $x_1 + \cdots + x_m$, while $i$'s price is non-negative.) Thus, because of the Bolzano-Weierstrass theorem (i.e., because any non-decreasing and upper bounded sequence of real numbers must converge), for every $s_{-i} \in S_{-i}$ and $t \in \{0, 1, \ldots, m\}$, letting $D \stackrel{\text{def}}{=} \min_{l \in [m]}\{x_l - y_l\}$, there must exist some $H_\varepsilon^{(s_{-i},t)} \in \mathbb{N}$ such that, for all $k > H_\varepsilon^{(s_{-i},t)}$:

$$\left| \Big(\sum_{j \in [m]} M_{i,j}^{\mathsf{A}}(\sigma_i^{(k)}, s_{-i})\big(\sum_{l=1}^j z_{t,l}\big) - M_i^{\mathsf{P}}(\sigma_i^{(k)}, s_{-i})\Big) \right.$$

$$\left. - \Big(\sum_{j \in [m]} M_{i,j}^{\mathsf{A}}(\widetilde{\sigma}_i^{(k)}, s_{-i})\big(\sum_{l=1}^j z_{t,l}\big) - M_i^{\mathsf{P}}(\widetilde{\sigma}_i^{(k)}, s_{-i})\Big) \right|$$

$$= \left| U_i\big(z_t, M(\sigma_i^{(k)}, s_{-i})\big) - U_i\big(z_t, M(\widetilde{\sigma}_i^{(k)}, s_{-i})\big) \right| \leq \frac{\varepsilon D}{4} \ . \tag{E.4}$$

At this point, we invoke the finiteness of the mechanism in order to define the following maximum value:

$$H_\varepsilon \stackrel{\text{def}}{=} \max\big\{ H_\varepsilon^{(s_{-i},t)} : s_{-i} \in S_{-i}, t \in \{0, 1, \ldots, m\} \big\} \in \mathbb{N} \ .$$

As a result, (E.4) holds for every $k > H_\varepsilon$, $s_{-i} \in S_{-i}$, and $t \in \{0, 1, \ldots, m\}$. We now claim that, by picking an arbitrary $k > H_\varepsilon$, the strategies $\sigma_i^{(k)}$ and $\widetilde{\sigma}_i^{(k)}$ must be the two 'sufficiently close' strategies we are looking for.

To prove this, consider an arbitrary strategy subprofile $s_{-i} \in S_{-i}$ and an integer $t \in [m]$, and apply (E.4) twice, once for $t$ and once for $t-1$. Combining the resulting two inequalities and applying the triangle inequality, we have:[19]

$$\frac{\varepsilon D}{2} \geq \Big| \ \Big(\sum_{j \in [m]} M_{i,j}^{\mathsf{A}}(\sigma_i^{(k)}, s_{-i})\big(\sum_{l=1}^j z_{t,l}\big) - M_i^{\mathsf{P}}(\sigma_i^{(k)}, s_{-i})\Big)$$

$$- \Big(\sum_{j \in [m]} M_{i,j}^{\mathsf{A}}(\widetilde{\sigma}_i^{(k)}, s_{-i})\big(\sum_{l=1}^j z_{t,l}\big) - M_i^{\mathsf{P}}(\widetilde{\sigma}_i^{(k)}, s_{-i})\Big)$$

$$- \Big(\sum_{j \in [m]} M_{i,j}^{\mathsf{A}}(\sigma_i^{(k)}, s_{-i})\big(\sum_{l=1}^j z_{t-1,l}\big) - M_i^{\mathsf{P}}(\sigma_i^{(k)}, s_{-i})\Big)$$

$$+ \Big(\sum_{j \in [m]} M_{i,j}^{\mathsf{A}}(\widetilde{\sigma}_i^{(k)}, s_{-i})\big(\sum_{l=1}^j z_{t-1,l}\big) - M_i^{\mathsf{P}}(\widetilde{\sigma}_i^{(k)}, s_{-i})\Big) \ \Big|$$

$$= \left| \Big(\sum_{j=t}^m M_{i,j}^{\mathsf{A}}(\sigma_i^{(k)}, s_{-i})(x_t - y_t)\Big) - \Big(\sum_{j=t}^m M_{i,j}^{\mathsf{A}}(\widetilde{\sigma}_i^{(k)}, s_{-i})(x_t - y_t)\Big) \right|$$

$$= (x_t - y_t) \left| \Big(\sum_{j=t}^m M_{i,j}^{\mathsf{A}}(\sigma_i^{(k)}, s_{-i})\Big) - \Big(\sum_{j=t}^m M_{i,j}^{\mathsf{A}}(\widetilde{\sigma}_i^{(k)}, s_{-i})\Big) \right| \ ,$$

which further implies, using $x_t - y_t \geq D > 0$, that

$$\left| \Big(\sum_{j=t}^m M_{i,j}^{\mathsf{A}}(\sigma_i^{(k)}, s_{-i})\Big) - \Big(\sum_{j=t}^m M_{i,j}^{\mathsf{A}}(\widetilde{\sigma}_i^{(k)}, s_{-i})\Big) \right| \leq \frac{\varepsilon}{2} \ . \tag{E.5}$$

Let us now use (E.5) to argue that the following set of inequalities hold:

$$\forall t \in [m] \quad \left| M_{i,t}^{\mathsf{A}}(\sigma_i^{(k)}, s_{-i}) - M_{i,t}^{\mathsf{A}}(\widetilde{\sigma}_i^{(k)}, s_{-i}) \right| \leq \varepsilon \ . \tag{E.6}$$

Indeed, for $t = m$, (E.6) can be derived by plugging $t = m$ into (E.5). Else, for each $t \in \{1, 2, \ldots, m-1\}$, we apply (E.5) twice, once for $t$ and once for $t+1$, and again combine the resulting inequalities with the triangle inequality to deduce (E.6).

---

[19]That is, $|a - b| \leq \varepsilon$ and $|c - d| \leq \varepsilon$ imply $|(a - b) - (c - d)| \leq 2\varepsilon$.

This completes the proof of Lemma E.1. □

### E.2 Deducing Theorem 3 from Lemma E.1

Because $\mathbb{K}$ is $\varepsilon$-basic, let $V$ and $W$ be the corresponding subsets of reals from Definition 5.1. Denote by $a, b \in V \cap W$ any two disjoint reals in $V \cap W$ such that $a > b$. For each player $i$, consider the following two $\delta$-approximate candidate sets

$$\widetilde{K}_i \stackrel{\text{def}}{=} \{\theta_i \in \Theta_i \mid \forall j, \theta_i(j) \in V\} \quad \text{and} \quad \widetilde{K}'_i \stackrel{\text{def}}{=} \{\theta_i \in \Theta_i \mid \forall j, \theta_i(j) \in W\} \ ,$$

and according to the $\varepsilon$-basic assumption on $\mathbb{K}_i$, we have $\widetilde{K}_i, \widetilde{K}'_i \in \mathbb{K}_i$. Next, consider the following two valuations that belong to $\Theta_i$ for every $i$:

$$x = (a, a, \ldots, a) \quad \text{and} \quad y = (b, b, \ldots, b) \ .$$

It is simple to verify that $x$, $y$, $\widetilde{K}_i$ and $\widetilde{K}'_i$ satisfy the hypothesis of Lemma E.1 (or more precisely, (E.1)). Thus, for any $\varepsilon' > 0$, the following holds:

$$\text{for all } i \in [n] \text{ there exist } \sigma_i \in \Delta(\mathsf{UD}_i(\widetilde{K}_i)) \text{ and } \sigma'_i \in \Delta(\mathsf{UD}_i(\widetilde{K}'_i)) \text{ such that} \quad (E.7)$$
$$\forall s_{-i} \in S_{-i} \ \forall j \in [m] \quad \left| M^{\mathsf{A}}_{i,j}(\sigma_i, s_{-i}) - M^{\mathsf{A}}_{i,j}(\sigma'_i, s_{-i}) \right| < \varepsilon' \ .$$

Consider the allocation of $M$ under the strategy profile $\sigma' = (\sigma'_1, \sigma'_2, \ldots, \sigma'_n)$. Because there are $m$ copies of the good, there ought to be one player who, in expectation, receives no more than $\frac{m}{n}$ copies. Without loss of generality, let him be player 1: that is, $\sum_{j=1}^{m} j \cdot M^{\mathsf{A}}_{1,j}(\sigma'_1, \ldots, \sigma'_n) \leq \frac{m}{n}$. Thus, by (E.7) and multiple applications of the triangle inequality, we have

$$\sum_{j=1}^{m} j \cdot M^{\mathsf{A}}_{1,j}(\sigma_1, \sigma'_{-1}) \leq \frac{m}{n} + \varepsilon' m^2 \ .$$

By averaging, there exists a *pure* strategy profile $s = (s_1, s_{-1})$ in the support of $(\sigma_1, \sigma'_{-1})$ satisfying

$$\sum_{j=1}^{m} j \cdot M^{\mathsf{A}}_{1,j}(s_1, s_{-1}) \leq \frac{m}{n} + \varepsilon' m^2 \ . \quad (E.8)$$

Now let

$$K \stackrel{\text{def}}{=} (K_1, \ldots, K_n), \text{ where } K_i \stackrel{\text{def}}{=} \begin{cases} \widetilde{K}_1 & \text{if } i = 1 \\ \widetilde{K}'_i & \text{if } i = 2, \ldots, n \end{cases}$$

$$\theta \stackrel{\text{def}}{=} (\theta_1, \ldots, \theta_n), \text{ where } \theta_i \stackrel{\text{def}}{=} \begin{cases} (\max V, \ldots, \max V) & \text{if } i = 1 \\ (\min W, \ldots, \min W) & \text{if } i = 2, \ldots, n. \end{cases}$$

Because we know that $(\sigma_1, \sigma'_{-1}) \in \Delta(\mathsf{UD}_1(K_1)) \times \cdots \times \Delta(\mathsf{UD}_n(K_n))$ from (E.7), we deduce that $s \in \mathsf{UD}(K)$. It is also obvious that $\theta \in K$ and $\mathrm{MSW}(\theta) = m \cdot \max V$.

Next, we show that $s$, $K$, and $\theta$ satisfy the desired inequality of Theorem 3. Indeed,

$$\mathbb{E}\Big[\mathrm{SW}\big(\theta, M(s)\big)\Big] \stackrel{(*)}{\leq} \Big(\frac{m}{n} + \varepsilon' m^2\Big) \cdot \max V + \Big(m - \frac{m}{n} - \varepsilon' m^2\Big) \cdot \min W$$

$$= m \max V - \Big(m - \frac{m}{n} - \varepsilon' m^2\Big)(\max V - \min W)$$

$$\leq m \max V - \Big(m - \frac{m}{n} - \varepsilon' m^2\Big)\Big(2\delta - \frac{\varepsilon}{m}\Big)$$

$$= \mathrm{MSW}(\theta) - 2\delta m\big(1 - 1/n\big) + 2\delta\varepsilon' m^2 + \frac{\varepsilon}{m}\Big(m - \frac{m}{n} - \varepsilon' m^2\Big)$$

$$\leq \mathrm{MSW}(\theta) - 2\delta m\big(1 - 1/n\big) + \varepsilon + 2\delta\varepsilon' m^2 - \frac{\varepsilon}{n} \ .$$

Above, inequality $(*)$ holds because, when $\theta$ is the true-valuation profile, the value for each copy of the good is $\max V$ for player 1, and is $\min W$ for every player other than player 1. However, in the outcome $M(s)$, owing to (E.8), in expectation player 1 can receive at most $\frac{m}{n} + \varepsilon' m^2$ copies of the good.

Finally, noticing that $\varepsilon' > 0$ can be arbitrarily small, we can choose $\varepsilon'$ to satisfy $2\delta\varepsilon' m^2 - \frac{\varepsilon}{n} \leq 0$. This implies that $\mathbb{E}[\mathrm{SW}(\theta, M(s))] \leq \mathrm{MSW}(\theta) - 2\delta m(1 - 1/n) + \varepsilon$. Therefore, Theorem 3 holds.
∎

## F  The Set of Undominated Strategies is Non-Empty

It is trivial to see that, no matter what candidate set $K_i$ a player $i$ may have, $\mathsf{UD}_i(K_i)$ is non-empty in the Vickrey mechanism. In fact, Theorem 2 implies that $\mathsf{UD}_i(K_i)$ includes at least all the valuations in $K_i$.

Below, we argue that $\mathsf{UD}_i(K_i)$ is also always non-empty for all finite mechanisms.

**Fact F.1.** *Let $M$ be a finite mechanism, $i$ a player, and $K_i$ a candidate set of $i$. Then, $\mathsf{UD}_i(K_i) \neq \varnothing$.*

*Proof.* Let $S_i = \{s_1, \ldots, s_t\}$ be the finite pure-strategy set of player $i$. We proceed by contradiction. Suppose that every strategy in $S_i$ is (weakly) dominated, with respect to $K_i$, by some strategy in $\Delta(S_i)$. Then, in particular, $s_1$ is dominated. Thus, there exists a mixed strategy $\sum_{k=1}^{t} \alpha_k s_k \in \Delta(S_i)$ such that

$$s_1 \prec_{(i,K_i)} \sum_{k=1}^{t} \alpha_k s_k \ , \tag{F.1}$$

where $\alpha \in \Delta \stackrel{\text{def}}{=} \{x \in [0,1]^t \mid \sum_{k=1}^{t} x_k = 1\}$. Notice that, by condition (2) in Definition 4.1, we cannot have $s_1 \prec_{(i,K_i)} s_1$. Therefore, we must have $\alpha_1 < 1$. Now, we simplify (F.1) by subtracting $\alpha_1 s_1$ on both sides and rescaling:

$$s_1 \prec_{(i,K_i)} \sum_{k=2}^{t} \frac{\alpha_k}{1-\alpha_1} s_k \ . \tag{F.2}$$

Next, since $s_2$ is dominated, let it be dominated by $\sum_{k=1}^{t} \beta_k s_k$. In symbols,

$$s_2 \prec_{(i,K_i)} \sum_{k=1}^{t} \beta_k s_k \tag{F.3}$$

for some $\beta \in \Delta$. By substituting (F.2) into (F.3), we can rewrite (F.3) as

$$s_2 \prec_{(i,K_i)} \sum_{k=2}^{t} \beta'_k s_k \tag{F.4}$$

for some $\beta' \in \Delta$ such that $\beta'_1 = 0$. Again, by subtracting $\beta'_2 s_2$ on both sides and rescaling, we obtain

$$s_2 \prec_{(i,K_i)} \sum_{k=3}^{t} \beta''_k s_k \ , \tag{F.5}$$

for some $\beta'' \in \Delta$ such that $\beta''_1 = \beta''_2 = 0$. We substitute (F.5) into (F.2), and obtain

$$s_1 \prec_{(i,K_i)} \sum_{k=3}^{t} \alpha'_k s_k \ ,$$

for some $\alpha' \in \Delta$ such that $\alpha'_1 = \alpha'_2 = 0$.

This process, similar to Gaussian elimination in linear systems, can be continued until we obtain $s_k \prec_{(i,K_i)} s_t$ for every $k = 1, \ldots, t-1$. Thus, $s_t$ must be an undominated strategy for player $i$, contradicting the hypothesis that $\mathsf{UD}_i(K_i) = \varnothing$. □

## G  The Work of Lopomo, Rigotti, and Shannon

**Their Model.**  In order to "strip away issues pertaining to higher order beliefs and strategic uncertainty", Lopomo, Rigotti, and Shannon [21] focus on single-player mechanisms. Thus, so do we when recalling their work.

In their model, *true state of the world* comprises all the information the player is uncertain about, and the player's utility function, $U$, maps $O \times T \times S$ to $\mathbb{R}$, where

(a) $O$ is the set of all possible outcomes,

(b) $T \stackrel{\text{def}}{=} [0,1]$ is the set of all possible player types, and

(c) $S$ is the set of all possible true states of the world.

When the player's type is $t \in T$, the only information the player has about the true state of the world $s \in S$ is that $s$ is drawn from a distribution $\Pi(t)$ in $\Delta(S)$.

In their model, the player knows his own type $t \in T$, and a mechanism knows the true state of the world $s \in S$. The player is allowed to report just his own type, and then a mechanism chooses an outcome based not only on this report, but also on the true state: that is, each mechanism $\phi$ is a function $\phi \colon T \times S \to O$.

By contrast, in our auction setting, a mechanism chooses an outcome solely based on the players' reports. Indeed, since each player is uncertain about his own valuation, the true state of the world should include the true valuation profile $\theta^*$, and if a mechanism knew $\theta^*$, then it would be trivial to choose an outcome of maximum social welfare.

In their Knightian setting, they provide a general notion of a dominant-strategy mechanism, *optimal incentive compatibility (optimal IC)*, and a very restrictive notion of a dominant-strategy mechanism, *ex-post incentive compatibility (ex-post IC)*. Formally, a mechanism $\phi$ is

- optimal IC if, $\forall t \in T$, $\forall \sigma \in \Delta(T)$, and $\forall \pi \in \Pi(t)$: $\mathbb{E}_{s \sim \pi}\big[U(\phi(t,s),t,s)\big] \geq \mathbb{E}_{s \sim \pi}\big[\mathbb{E}_{\theta \sim \sigma}\big[U(\phi(\theta,s),t,s)\big]\big]$, and

- ex-post IC if, $\forall t, \theta \in T$ and $\forall s \in S$: $U(\phi(t,s),t,s) \geq U(\phi(\theta,s),t,s)$.

**Their First Theorem.**  They assume that, for every type $t \in T$, there exists a neighborhood $N(t) \subset T$ such that, for all continuous functions $g \colon S \to \mathbb{R}$,

$$\text{if } \int_S g(s) d\pi = 0 \text{ for every } \pi \in \bigcap_{t' \in N(t)} \Pi(t'), \text{ then } g = 0.$$

Under this assumption, their first theorem shows that every optimal IC mechanism satisfying an additional technical condition (i.e., ex-post cyclical monotonicity) must be ex-post IC.

Therefore, their first theorem has the same spirit of our Theorem 1. In both theorems, some form of overlapping of a player's possible 'belief/knowledge sets' implies that every dominant-strategy mechanism must be of a very restrictive form. However, due to the differences in models and assumptions, it is unclear whether our already simple proof of Theorem 1 can be more simply derived from theirs. Even ignoring all other differences, there cannot be any subjective map from their type space to ours. In their case, a player's type space (i.e., $T = [0,1]$) has the cardinality of the continuum. In our case, the type space of a given player $i$ (i.e., $\mathbb{K}_i$) may have the cardinality of the power set of the continuum.

[17] Matthew O. Jackson. Implementation in undominated strategies: A look at bounded mechanisms. *Review of Economic Studies*, 59(4):757–75, October 1992.

[18] Matthew O. Jackson, Thomas Palfrey, and Sanjay Srivastava. Undominated Nash implementation in bounded mechanisms. *Games and Economic Behavior*, 6(3):474–501, 1994.

[19] Frank H. Knight. *Risk, Uncertainty and Profit*. Houghton Mifflin, 1921.

[20] Kevin Leyton-Brown and Yoav Shoham. Essentials of game theory: A concise multidisciplinary introduction. *Synthesis Lectures on Artificial Intelligence and Machine Learning*, 2(1):1–88, 2008.

[21] Giuseppe Lopomo, Luca Rigotti, and Chris Shannon. Uncertainty in mechanism design. Technical report, 2009.

[22] Giuseppe Lopomo, Luca Rigotti, and Chris Shannon. Knightian uncertainty and moral hazard. *Journal of Economic Theory*, 146(3):1148 – 1172, 2011. Incompleteness and Uncertainty in Economics.

[23] Fabio Maccheroni, Massimo Marinacci, and Aldo Rustichini. Ambiguity aversion, robustness, and the variational representation of preferences. *Econometrica*, 74(6):1447–1498, 2006.

[24] Andrew Mas-Colell. An equilibrium existence theorem without complete or transitive preferences. *Journal of Mathematical Economics*, 1(3):237–246, December 1974.

[25] Roger B. Myerson. Incentive compatibility and the bargaining problem. *Econometrica: journal of the Econometric Society*, pages 61–73, 1979.

[26] Leandro Nascimento. Remarks on the consumer problem under incomplete preferences. *Theory and Decision*, 70(1):95–110, January 2011.

[27] Efe A. Ok. Utility representation of an incomplete preference relation. *Journal of Economic Theory*, 104:429–449, 2002.

[28] Luca Rigotti and Chris Shannon. Uncertainty and risk in financial markets. *Econometrica*, 73(1):203–243, 01 2005.

[29] David Schmeidler. Subjective probability and expected utility without additivity. *Econometrica*, 57(3):571–87, May 1989.

[30] Wayne Shafer and Hugo Sonnenschein. Equilibrium in abstract economies without ordered preferences. *Journal of Mathematical Economics*, 2(3):345–348, December 1975.

[31] William Vickrey. Counterspeculation, auctions, and competitive sealed tenders. *The Journal of Finance*, 16(1):8–37, 1961.